\documentclass[twocolumn,twoside,slac_two]{revtex4}
\usepackage{graphicx}
\usepackage{fancyhdr}
\usepackage{xspace}
\newcommand{\sqrts}{$\sqrt{s}$\xspace}
\newcommand{\pT}{$p_{T}$\xspace}
\newcommand{\pp}{{\it p-p}\xspace}

\begin{document}

\title{Underlying Event Studies at RHIC}

%

\author{Helen Caines for the STAR Collaboration}
\affiliation{Department of Physics, Yale  University, New Haven, CT 06520, USA}

\begin{abstract}
By studying p-p collisions we hope to improve our understanding of the fundamental constituents of matter and how they form into colorless objects. Measurements of the inclusive jet cross-sections and fragmentation properties have confirmed that QCD based calculations give a good description of the hard scattering processes. However, as our analysis of jets has improved it has become clear that there is significant contribution to these measurements from processes other than those directly related to the initial hard scattering - the so-called underlying event. Several processes contribute to the underlying event, namely the beam-beam remnants,  initial and final state radiation and multiple parton interactions. The structure of the jet and the underlying event are strikingly different in both their particle compositions and momentum distributions. Only by understanding both components can one fully describe a p-p collision. I will discuss preliminary results from studies of the underlying event in p-p collisions at $\sqrt{s}$ = 200 GeV at RHIC, and compare to PYTHIA predictions, as well as earlier results from the Tevatron at 1.8-1.96 TeV.
\end{abstract}

\maketitle

\thispagestyle{fancy}



\section{Introduction}

The study of the properties of jets  and the underlying event in \pp collisions is important for improving our understanding of QCD and the hadronization process, as well as providing a vital baseline for comparison to measurements being performed at RHIC by the STAR and PHENIX collaborations in heavy-ion collisions~\cite{QMBruna, QMPloskon, QMLai, QMSalur}.  

The results presented here are a preliminary study of  \pp collisions at \sqrts  = 200 GeV by the STAR collaboration.  It has previously been shown~\cite{STARppJetXsec} that the inclusive jet cross-section in \pp collisions at  \sqrts = 200 GeV is well described over seven orders of magnitude by NLO pQCD calculations. Further, these theoretical calculations can also successfully describe the $\pi$ and (anti)proton production in minimum bias collisions at the same collision energy~\cite{ppPi, ppP}.  These results indicate that we have a well calibrated probe. However, the fragmentation functions, jet shapes, particle composition of the hadronized fragments and the underlying event all remain unexplored at RHIC energies.  Such studies are now underway and a few selected preliminary results are presented below.  

Calculations suggest that the identified particle fragmentation functions are modified  in different ways by the presence of a Quark Gluon Plasma~\cite{Sapeta07}, this makes it especially important to have a well measured baseline with which to compare to the measurements in Au-Au collisions.

\begin{figure}[htb]
	\begin{center}
		\includegraphics[width=\linewidth]{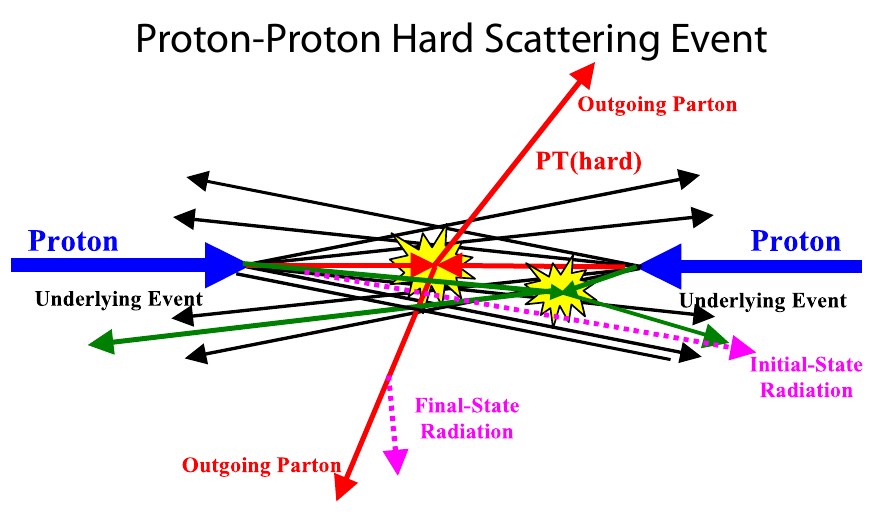}
		\caption{ Cartoon illustrating the major components used in QCD Monte-Carlo models to  simulate a  2-to-2 parton hard scattering in a  proton-proton collision.  The event consists of two outgoing jets from the hard scattered partons and the underlying event. The jets contain particles from the fragmentation of the initial partons  plus initial and final state radiation. The protons break up into beam-beam remnants. PYTHIA also includes the possibility of the second semi-hard 2-to-2 parton-parton scattering (multiple parton interactions). The underlying event consists of the beam-beam remnants, hadrons resulting from the multiple parton interaction and the initial and final state radiation hadrons not clustered in the hard scattered jets. Figure adapted from ~\cite{UEField}.}
		\label{Fig:ppCollision}
	\end{center}
\end{figure}

These data are also compared to predictions of  PYTHIA 6.410~\cite{Pythia}, tuned to the CDF 1.8-1.96 TeV data (Tune A). This  tuning has been shown to give good agreement with most  of the measurements made at the Tevatron~\cite{PythiaTuneA}. An important test of whether such a tuning is  truly representative of the underlying physics in elementary collisions, or merely a diligent selection of numerous parameters, is to compare PYTHIA predictions at a different collision energy without changing the parameters except via the estimated energy scaling.  A cartoon of many of the major components included in PYTHIA's modeling of  an elementary collision is shown in Fig.~\ref{Fig:ppCollision}. One such  important energy scaling parameter in PYTHIA is the hard scattering cut-off of multiple parton interactions (MPI) in the collision.  This cut-off is placed at p$_{T0}$(E$_{cm}$) = p$_{T0}$(E$_{cm}$/E$_{0}$)$^{\epsilon}$, where E$_{0}$ is 1.8-1.96 TeV, the tuning collision energy.  The initial estimate of $\epsilon$= 0.16 has been widely used in LHC predictions, however improved data at \sqrts = 1.8-1.96 TeV and newer data at \sqrts = 630 GeV now suggest that a value of $\epsilon$= 0.25 is more appropriate~\cite{PythiaTuneA}.  The inclusion of multiple parton interactions increases the activity, and hence particle production, in the underlying event. Increasing $\epsilon$ creates a larger energy dependence of the multiple parton interactions. A change in PYTHIA of $\epsilon$ from 0.16 to 0.25 results in a 35$\%$ increase in the charged   particle density in the underlying event at RHIC energies, while decreasing the particle density at the LHC by 26$\%$. It is therefore important to determine the correct scaling parameter in order to have accurate predictions for the LHC when it begins collisions later this year.

\section{The Analysis}

We utilize data from the mid-rapidity Time Projection Chamber (TPC) to reconstruct charged particles, and the mid-rapidity Barrel Electromagnetic Calorimeter (BEMC) to measure the neutral particle production. A "jet patch" trigger requiring  E$_{T}>$8 GeV in a  $\Delta \eta$ x $\Delta \phi$ = 1x1 patch of the BEMC was used to collect the data.  During the RHIC Run-6 in 2006 the use of this trigger allowed STAR to sample a luminosity of $\sim$8.7 pb$^{-1}$. By requiring such a large signal in the  BEMC   a neutral energy fragmentation bias is created for the triggering jet. Charged particle fragmentation functions are therefore presented only  for the di-jet partner not associated with the triggered jet-patch. 

\begin{figure}[htb]
	\begin{center}
		\includegraphics[width=\linewidth]{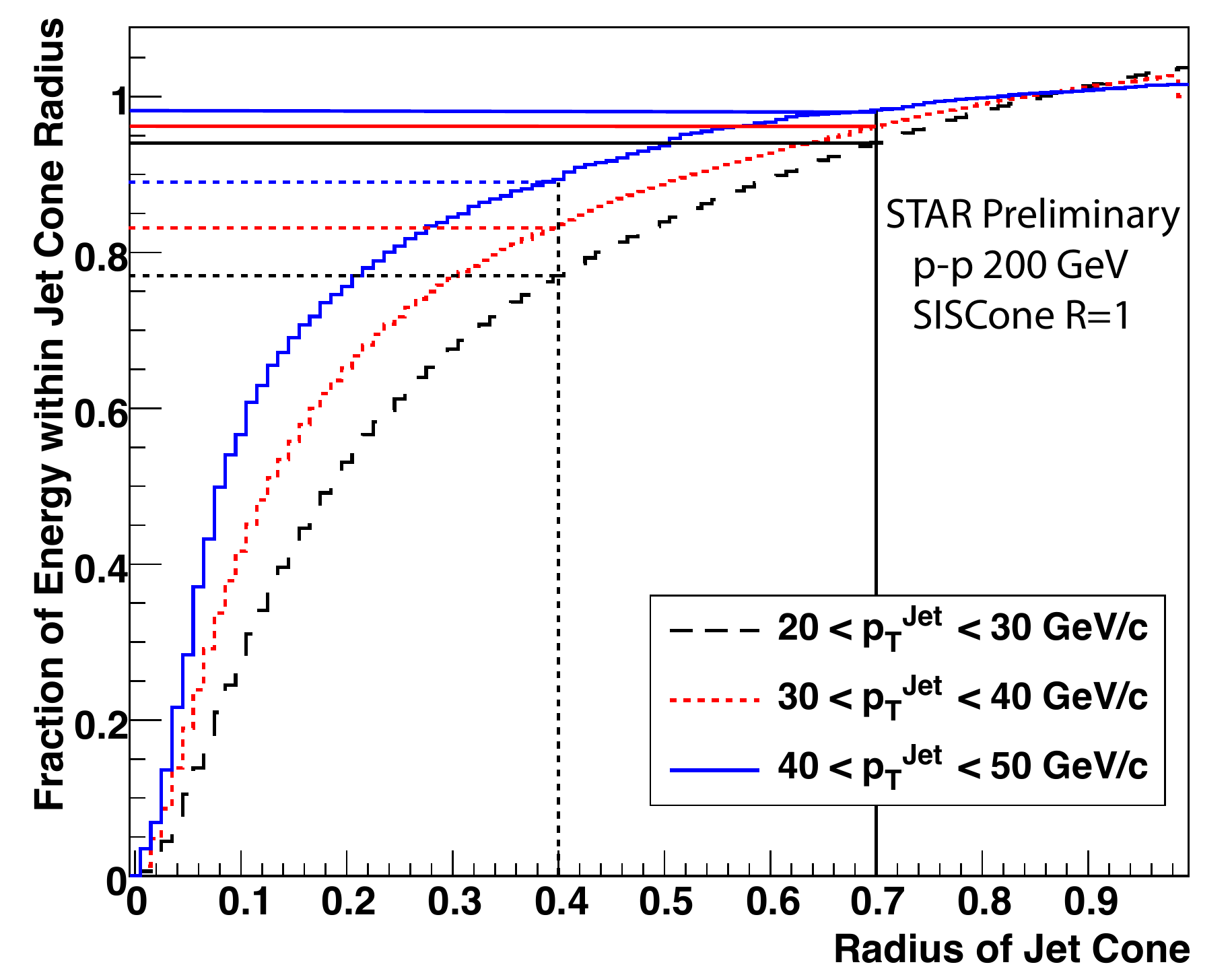}
		\caption{ Color online: The fraction of the jet's energy contained within jet cone radius R. Jets were initially found using the SISCone algorithm with R=1. }
		\label{Fig:JetRadius}
	\end{center}
\end{figure}

The k$_{T}$ and anti-k$_{T}$  and SISCone jet algorithms from the FastJet package~\cite{fastjet} were used to reconstruct jets, all three algorithms combine particles in  $\eta-\phi$ space and are infrared and collinear safe.  SISCone, a seedless cone algorithm, merges particles within a cone radius R=$\sqrt{\Delta\phi^2+\Delta\eta^2}$. A process of splitting and merging of cones is then performed to improve on the effectiveness of the cone algorithm in collecting  energy radiated off of the scattered parton via higher-order QCD processes.  The k$_{T}$ and anti-k$_{T}$  jet finders are sequential recombination algorithms. Such algorithms successively cluster in a  pair-wise fashion, all particles and proto-jets (previously merged particle pairs) in the event  via a \pT and distance weighted cut selection  until no cluster pairs pass the cut.  All remaining clusters are designated jets. For the k$_{T}$ algorithm the selection starts from the lowest  d$_{ij}$ particle/cluster pair, where  $d_{ij} = min( k_{Ti}^2, k_{Tj}^2) ( \Delta\phi_{ij}^{2}+\Delta\eta_{ij})^2)/R^2$, R is the resolution parameter and k$_{T}$ == \pT. If this minimum distance $d_{ij} <  k_{Ti}^2$, the pair are merged, and the iteration over all the new d$_{ij}$ continues until the test fails.  The anti-k$_{T}$ algorithm is similar except it merges from the highest d$_{ij}$ and selects for  $d_{ij} = min( 1/k_{Ti}^2, 1/k_{Tj}^2) ( \Delta\phi_{ij}^2+\Delta\eta_{ij}^2)/R^2$. In this way the anti-k$_{T}$ algorithm behaves much like an idealized cone algorithm.  A cut of $p_T$$>$0.2 GeV/c was applied to all charged particles considered in the event, and $E_{T} $$>$0.2 GeV for each tower cluster reconstructed in the BEMC. To investigate how the radius/resolution parameters, R,  used in this study affect the reconstructed jet energy, the SISCone algorithm was first run with R=1. The energy contained within this jet cone as a function of R was then studied, Fig.~\ref{Fig:JetRadius},  it can be seen that higher energy jets are focussed within smaller jet radii. For jets with 20 GeV/c $< p_{T}^{jet}<$ 30 GeV/c  $>$75 (95)$\%$ of the jet's energy is contained within R=0.4 (0.7) , while $>$89 (98)$\%$ of the jet's energy is contained within R=0.4 (0.7) for  40 GeV/c $< p_{T}^{jet} <$ 50 GeV/c.

The data are not yet corrected to the particle level, therefore the results are  compared to PYTHIA 6.410 calculations passed through STAR's detector simulation and reconstruction algorithms. The single charged  particle reconstruction efficiency in the TPC is $>$ 80$\%$ for \pT$>$1 GeV/c. Such detector inefficiencies  and  the presence of  particles which are undetected by the STAR detector, such as the neutron and K$^{0}_{L}$, cause the  reconstructed jet \pT  to be lower on average than the  true value.  The errors shown in all plots are therefore statistical only at this point.

\begin{figure}[htb]
	\begin{center}
		\includegraphics[width=\linewidth]{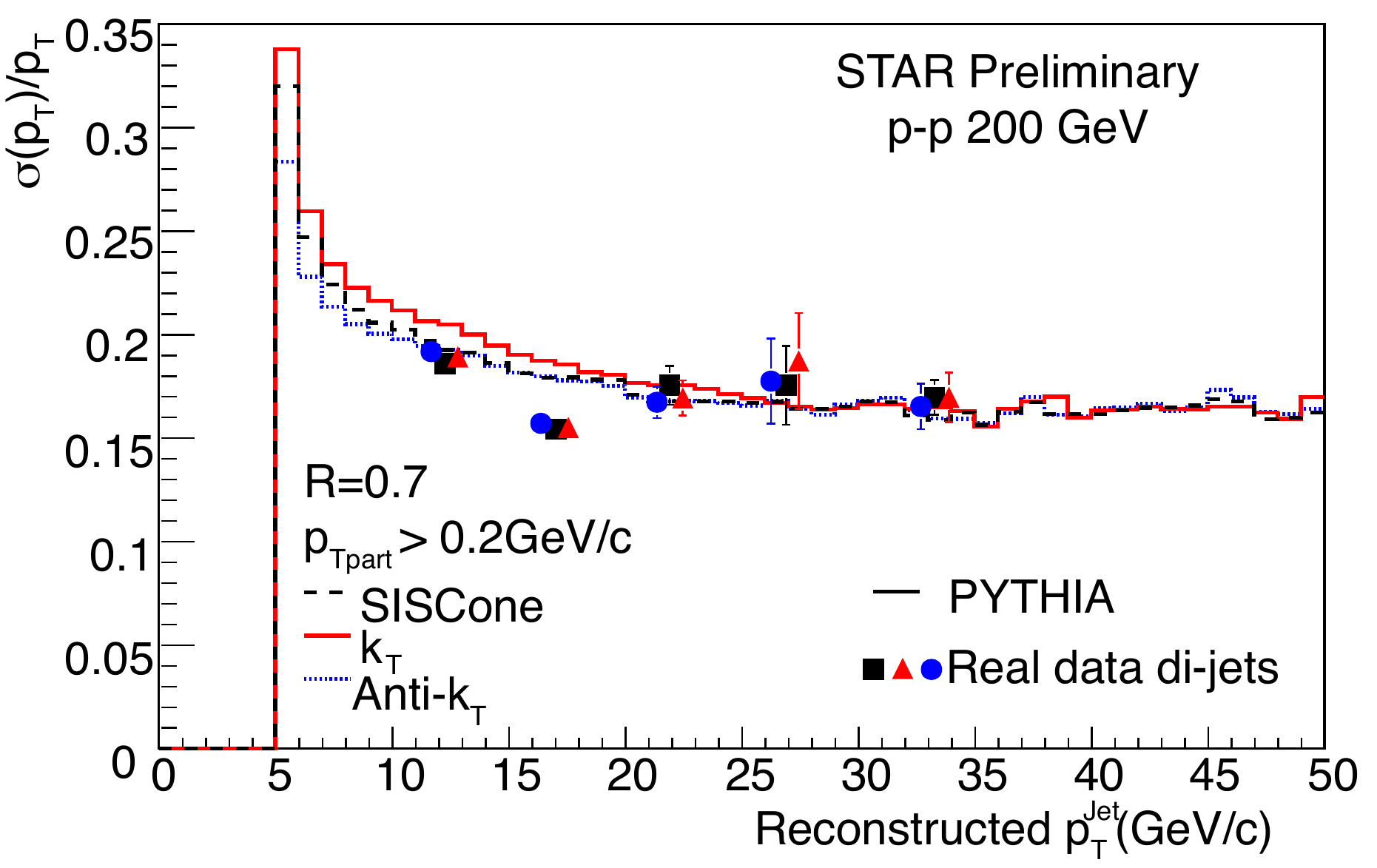}
	\caption{ Color online: The reconstructed jet energy resolution determined from PYTHIA simulations (histograms) and real di-jet data for the three jet algorithms used,  $|\eta|<$1-R, R=0.7. Red triangles - k$_{T}$,  blue circles - anti-k$_{T}$ and SISCone - black squares.} 	\label{Fig:JetRes}
		\end{center}
\end{figure}

The jet energy resolution was  obtained via two techniques. The first uses the PYTHIA simulations to compare reconstructed jet energies at the particle and detector level. The second studies  the energy balance of ``back-to-back" di-jets in the real \pp data. Figure~\ref{Fig:JetRes} shows that both methods resulted in  comparable jet energy resolutions of  $\sim$20$\%$ for reconstructed jet \pT $>$ 10 GeV/c.
	
\section{Jets in \pp}

\begin{figure}[htb] 
	\begin{center}
		\includegraphics[width=\linewidth]{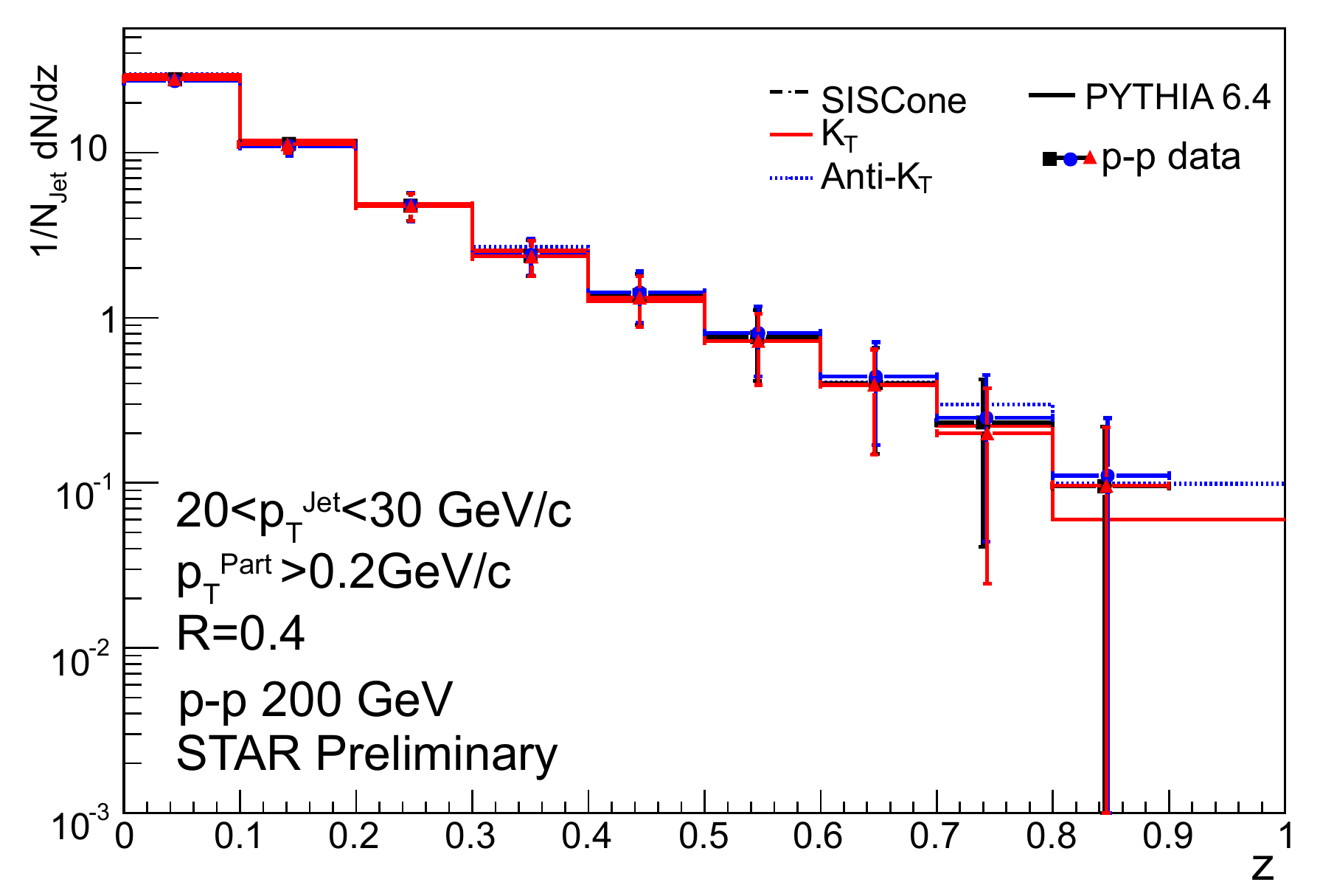}
		\caption{ Color online: Charged particle, detector level, $z$ fragmentation functions for jets reconstructed with 20$<$ \pT$<$ 30 GeV/c compared to PYTHIA for 3 different jet algorithms. $|\eta|<$1-R, R=0.4. Red triangles - k$_{T}$,  blue circles - anti-k$_{T}$ and SISCone - black squares.} \label{Fig:FFZ04}
	\end{center}
\end{figure}

\begin{figure}[htb] 
	\begin{center}
		\includegraphics[width=\linewidth]{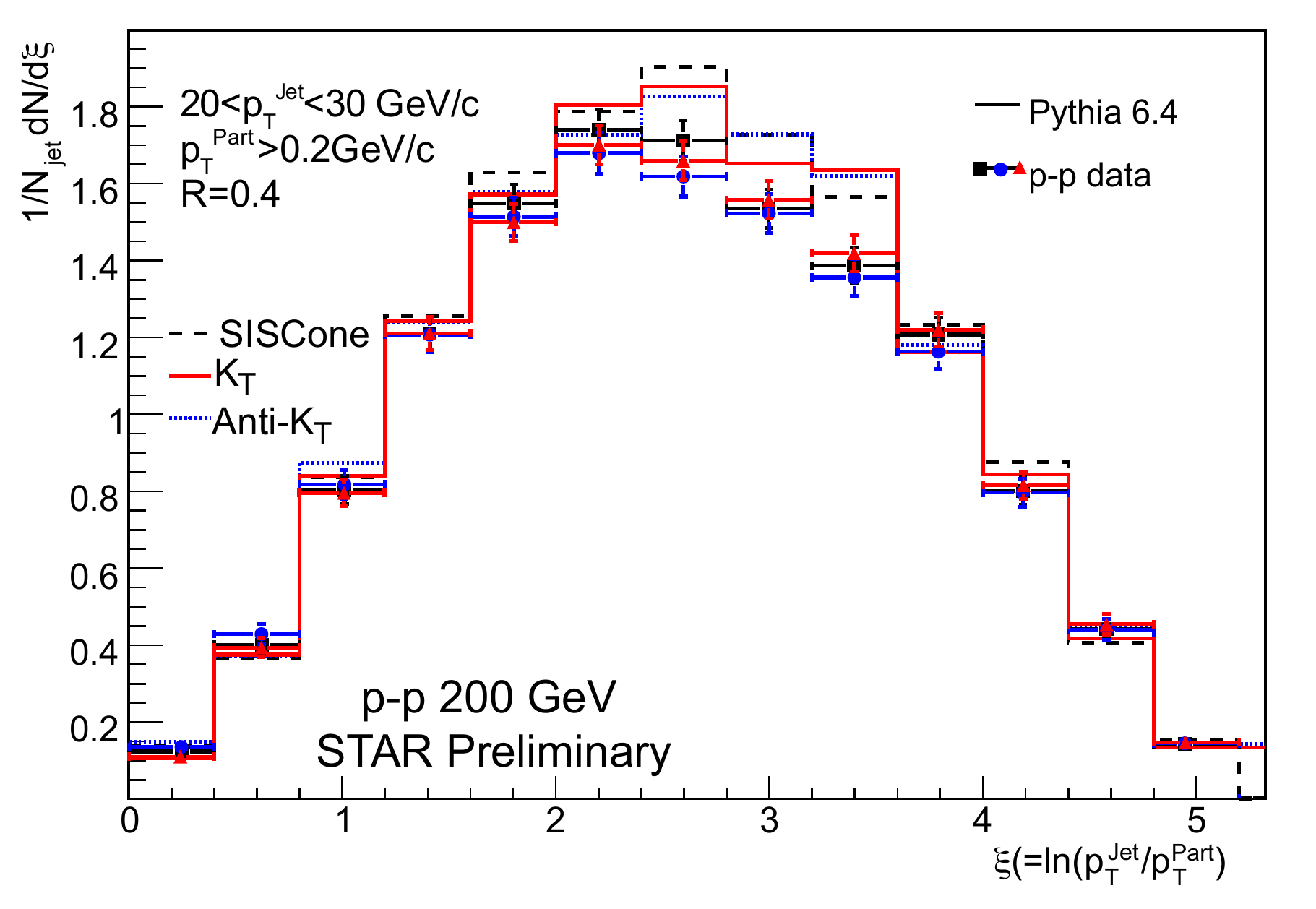}
	\caption{ Color online: Charged particle, detector level, $\xi$ fragmentation functions for jets reconstructed with 20$<$ \pT$<$ 30 GeV/c compared to PYTHIA for 3 different jet algorithms. $|\eta|<$1-R, R=0.4. Red triangles - k$_{T}$,  blue circles - anti-k$_{T}$  and SISCone - black squares.} \label{Fig:FFXi04}
		\end{center}		
\end{figure}

\begin{figure}[htb]
	\begin{center}
		\includegraphics[width=\linewidth]{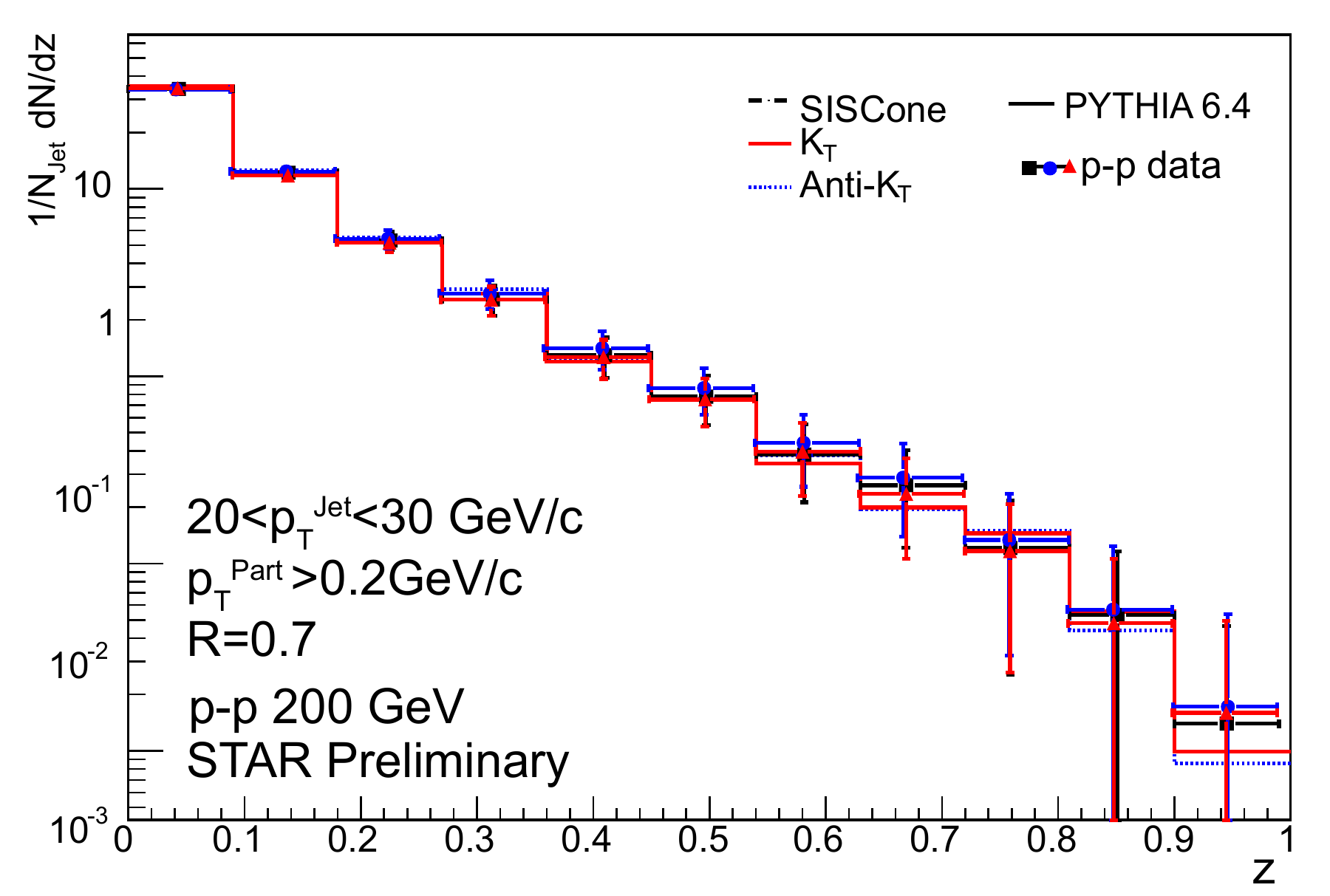}
		\caption{ Color online: Charged particle, detector level, $z$ fragmentation functions for jets reconstructed with 20$<$ \pT$<$ 30 GeV/c compared to PYTHIA for 3 different jet algorithms. $|\eta|<$1-R, R=0.7. Red triangles - k$_{T}$,  blue circles - anti-k$_{T}$  and SISCone - black squares.} \label{Fig:FFZ07}
	\end{center}
	
\end{figure}

\begin{figure}[htb]
	\begin{center}
		\includegraphics[width=\linewidth]{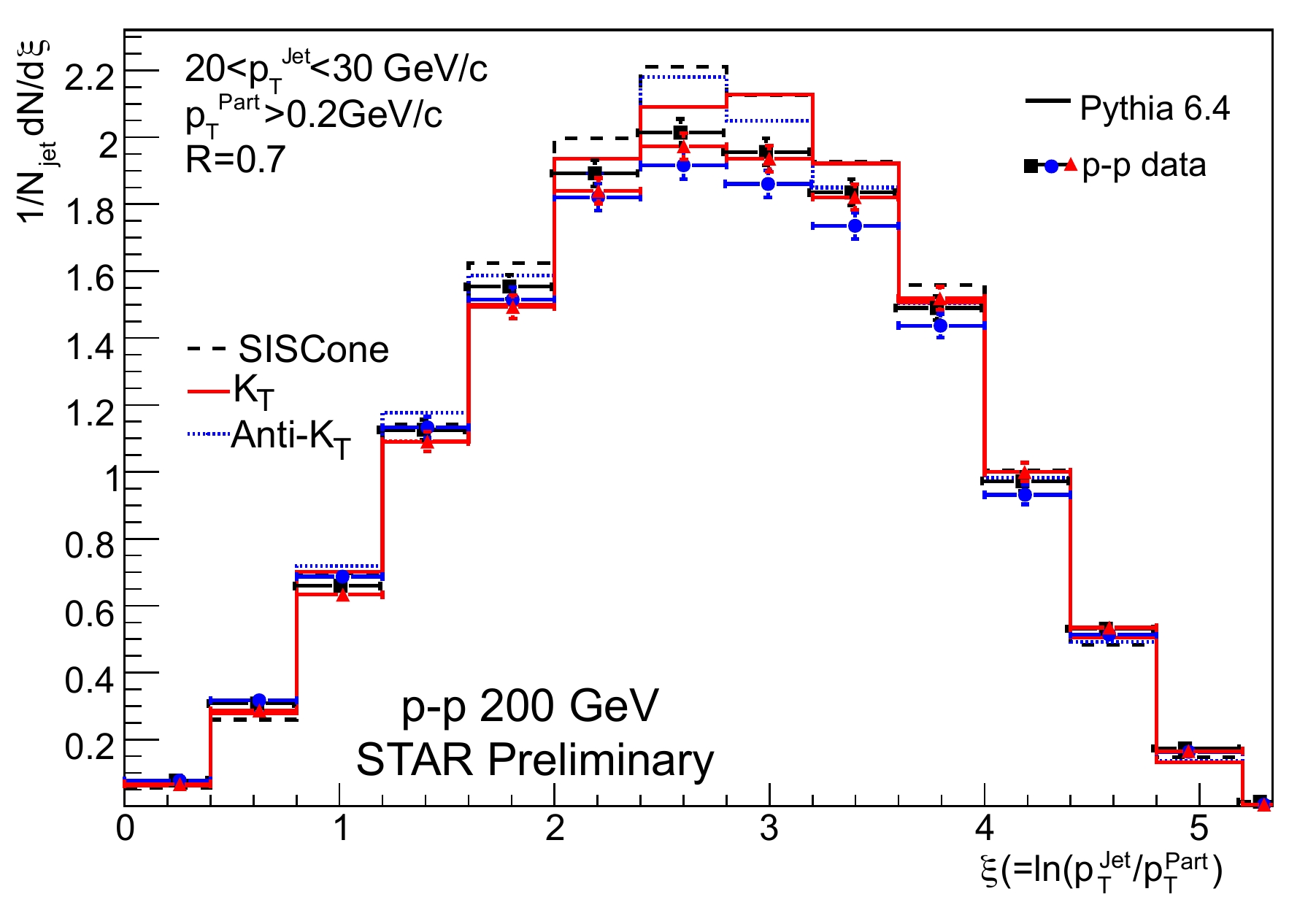}
	\caption{ Color online: Charged particle, detector level, $\xi$ fragmentation functions for jets reconstructed with 20$<$ \pT$<$ 30 GeV/c compared to PYTHIA for 3 different jet algorithms. $|\eta|<$1-R, R=0.7. Red triangles - k$_{T}$,  blue circles - anti-k$_{T}$  and SISCone - black squares.} \label{Fig:FFXi07}
		\end{center}
\end{figure}

 The uncorrected charged particle fragmentation functions  for jets with \pT reconstructed in the range 20-30 GeV/c are shown  for jet resolution parameters R=0.4, Fig.~\ref{Fig:FFZ04} and Fig.~\ref{Fig:FFXi04} and R=0.7, Fig.~\ref{Fig:FFZ07} and Fig.~\ref{Fig:FFXi07}. 
 
Figure~\ref{Fig:FFZ04} and  Fig.~\ref{Fig:FFZ07}  are the fragmentation functions as a function of z (=$p_T^{hadron}$/$p_T^{jet}$)  and Fig.~\ref{Fig:FFXi04} and  Fig.~\ref{Fig:FFXi07} as a function of $\xi$(=ln(1/z)). The solid points are the data and the histograms are the PYTHIA simulations. There is reasonable agreement between the data and PYTHIA, and the different jet algorithms  (shown as different colors and line types/shapes in the figures) reconstruct the same fragmentation functions within errors. This agreement, especially for the larger resolution parameter, suggests that at RHIC energies there are only minor NLO contributions beyond those approximated in the PYTHIA LO calculations.

\section{The Underlying Event}

\begin{figure}[ht]
		\begin{center}
			\includegraphics[width=\linewidth]{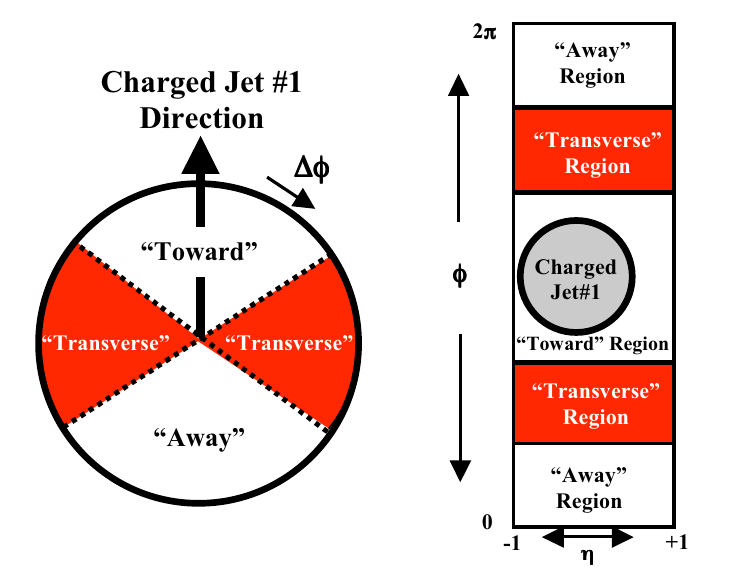}
			\caption{Cartoon of the areas in  $\Delta\phi$, relative to the leading jet, used to define the near jet , away jet , TransMin and TransMax regions for the underlying event analysis. Figure taken from ~\cite{UEField}.}
			\label{Fig:UEFig}
		\end{center}
\end{figure}

The underlying event   is defined as everything but the hard scattering. Thus, it has contributions from soft and semi-hard multiple parton interactions, initial and final state radiation and beam-beam remnants, Fig.~\ref{Fig:ppCollision}. Pile-up (multiple events being recorded as one by the experiment) is not considered to be part of the underlying event. This study is performed at mid-rapidity, hence the beam-beam contribution is minimal.  We follow the CDF technique~\cite{CDF} to perform this study. Briefly,  the method follows the following steps:  First the jets in an event are reconstructed, using the  algorithms mentioned above.  Next, the event is split into four sections defined by their azimuthal  angle with respect to the leading jet axis ($\Delta{\phi}$). The range within $|\Delta{\phi}|$$<$60$^{0}$ is the lead jet region. An away jet area is designated for $|\Delta{\phi}|$$>$120$^{0}$. This leaves two remaining transverse sectors of $60^{0}$$<$$\Delta{\phi}$$<$120$^{0}$ and $-120^{0}$$<$$\Delta{\phi}$$<$-60$^{0}$, see Fig.~\ref{Fig:UEFig}.  The transverse region containing the largest charged particle multiplicity  is assigned  the TransMax region, the other transverse sector is termed the TransMin region.  The  probability of the TransMax region containing contributions from initial and final state radiation components of the hard scattered parton is enhanced via this selection criteria. A ``leading" jet study, where at least one jet is found in STAR's acceptance,  and a ``back-to-back"  jet study, which is a sub-set of the ``leading" jet collection are performed.  This "back-to-back" sub-set of events  has two (and only two) found jets  with $p_{T}^{awayjet}/p_{T}^{leadjet}$$>$0.7 and $|\Delta{\phi_{jet}}|$$>$150$^{0}$. This selection suppresses the probability that the scattered parton undergoes any significant hard initial and final state radiation.  Thus, by comparing the TransMax and TransMin regions in the ``leading" and ``back-to-back" sets we can alter the sensitivity of the data  to the various components in the underlying event.

\begin{figure}[ht]
		\begin{center}
			\includegraphics[width=\linewidth]{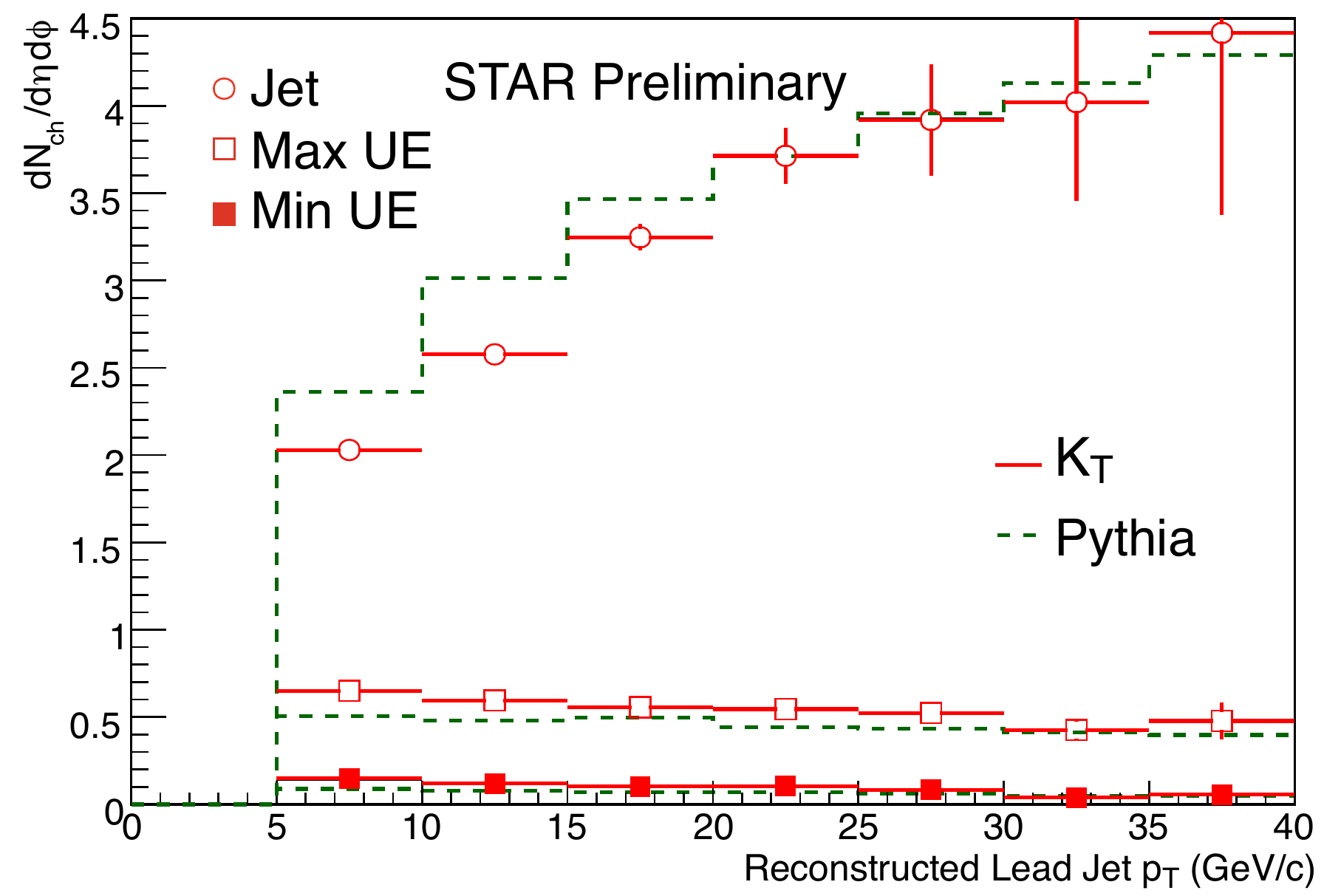}
			\caption{Color online: The uncorrected charged particle density in the ``back-to-back" data set for the away jet,  the TransMin and the TransMax regions as a function of reconstructed lead jet \pT, using k$_{T}$  algorithm, R=0.7. The dashed histograms indicate the predictions from PYTHIA. }
			\label{Fig:UENtracks}
		\end{center}
\end{figure}

\begin{figure}[ht]
		\begin{center}
			\includegraphics[width=\linewidth]{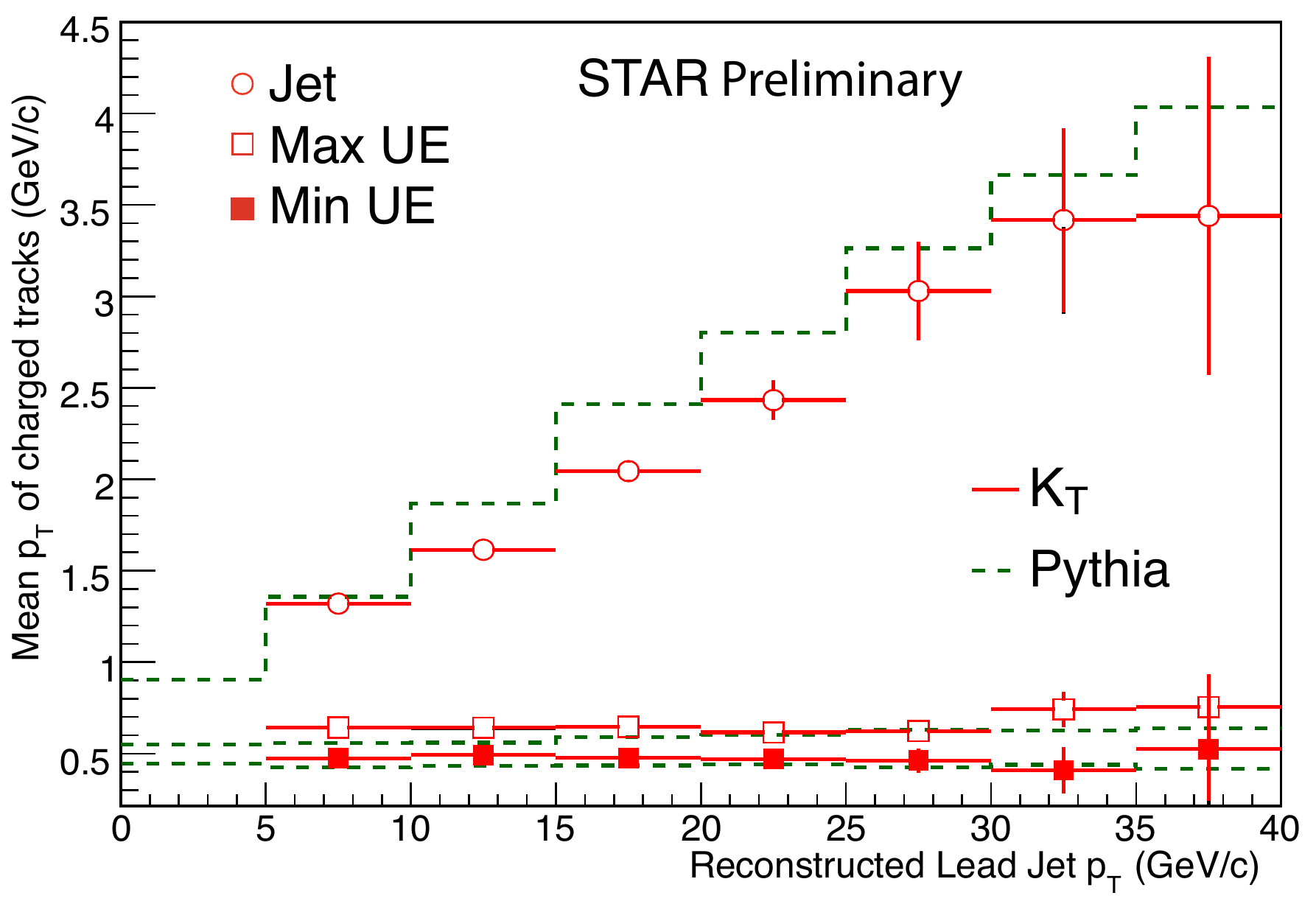}
			\caption{Color online: The uncorrected charged particle mean p$_{T}$  in the ``back-to-back" data set for the away jet,  the TransMin and the TransMax regions as a function of reconstructed lead jet \pT, using k$_{T}$  algorithm, R=0.7. The dashed histograms indicate the predictions from PYTHIA. }
			\label{Fig:UEPt}
		\end{center}
\end{figure}

Figure~\ref{Fig:UENtracks} shows the uncorrected charged particle densities of the away jet, TransMin and TransMax regions of the ``back-to-back" data set using the  k$_{T}$ algorithm, R=0.7. The first observation is that the underlying is largely independent of the jet energy while the away jet particle density steadily rises with reconstructed jet \pT, as expected. Figure~\ref{Fig:UEPt} shows the uncorrected mean \pT of these regions. Again the underlying event results are essentially independent of the hard scattering energy. The data for the underlying event at RHIC energies are slightly lower than those observed for the underlying event at CDF~\cite{CDF} . The results are consistent within errors for the SISCone, k$_{T}$, and anti-k$_{T}$ algorithms. In both figures the green dashed histogram is the prediction from PYTHIA. This PYTHIA tuning gets the trends of the data essentially correct, but slightly under predicts  the particle multiplicity and mean \pT in the Transverse regions, while over predicting the activity within the jet. In this tuning, as stated above, the multiple parton scattering energy scaling factor, $\epsilon$, is set to 0.25. The data is still slightly above PYTHIA  supporting the evidence that a  large value of the energy dependence parameter, $\epsilon$, is needed than that initially determined of $\epsilon$=0.16. This also suggests that a number of the LHC predictions, often labeled as ``ATLAS" tunes, are {\em over} predicting the activity in the underlying event at 14 TeV.

\begin{figure}[ht]
		\begin{center}
			\includegraphics[width=\linewidth]{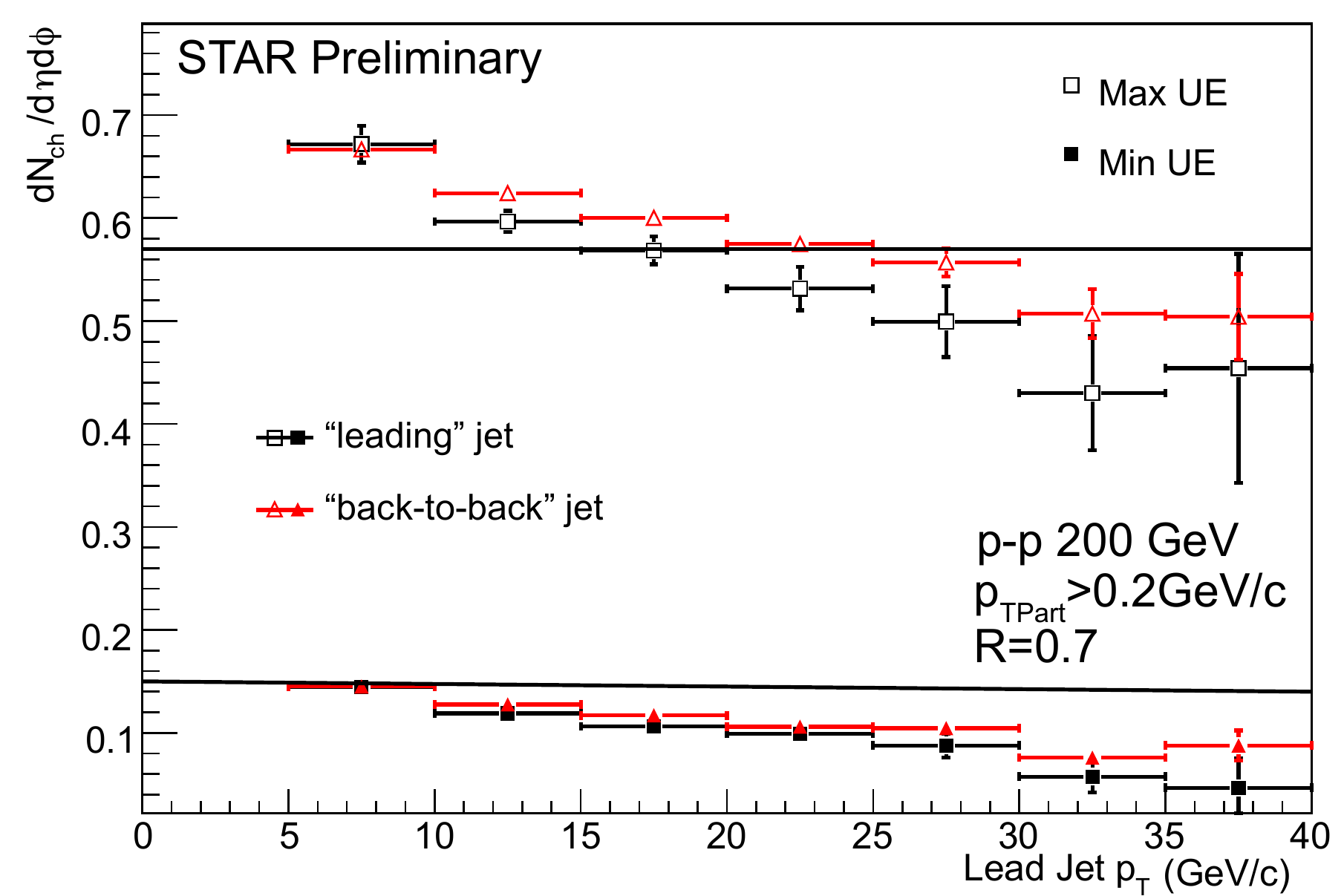}
			\caption{Color online: The uncorrected charged particle density in the TransMin and TransMax regions as a function of reconstructed lead jet \pT, using SISCone algorithm, R=0.7.  }
			\label{Fig:UEZoom}
		\end{center}

\end{figure}

Figure~\ref{Fig:UEZoom} shows the measured charged particle density in the underlying event. The densities are the same within errors for the ``leading" and ``back-to-back" datasets. This measurement, like those shown in Fig.~\ref{Fig:FFZ04}-\ref{Fig:FFXi07},  can be used to argue that the hard scattered partons emit  very small amounts of large angle initial/final state radiation at RHIC energies. This is very different in 1.96 TeV collisions where the ``leading"/``back-to-back" density ratio is $\sim$0.65~\cite{CDF}.  The two solid lines show the expected density if  events follow a Poisson distribution with an average of 0.36. The similarity of this simple simulation to the data suggests that at RHIC energies the splitting of the measured TransMax and TransMin values  is predominantly due to the sampling. PYTHIA results, not shown here, again display satisfactory agreement with the data for both data sets.

\section{Summary}

In summary, jet fragmentation functions have been measured in \pp collisions at \sqrts  = 200 GeV and will provide a stringent baseline for the measurements underway in Au-Au collisions. The three jet algorithms studied, SISCone, k$_{T}$ and anti-k$_{T}$ provide consistent results. PYTHIA, tuned to 1.8-1.96 TeV data, shows reasonable agreement suggesting that the the underlying physics and its energy dependence is well modeled. 

The underlying event is largely decoupled from magnitude of the momentum transfer of hard scattering. The data show there is virtually no large angle  initial and final state radiation from the hard scattering at RHIC energies. The collision energy dependence of the  multiple parton interactions in the event is more significant than initially estimated. This results in a smaller prediction for the magnitude of the  underlying event contribution to the background of the reconstructed jets at LHC energies. At RHIC the \pT spectra of particles in the underlying event are significantly softer than those from  jets. 

In the future these data will be corrected for detector inefficiencies and irresolution allowing for comparison to theoretical models at the particle level. The structure of the jet shapes (k$_{T}$, j$_{T}$ etc) will be studied, as will
the particle type composition of the jets and the underlying event. Ultimately  all these results will be compared to similar measurements made in heavy-ion collisions at the same beam energy.

\bigskip 

\end{document}